# Graphene/liquid crystal based terahertz phase shifters

Yang Wu,[1] Xuezhong Ruan,[1] Chih-Hsin Chen,[2] Young Jun Shin,[1] Youngbin Lee,[3] Jing Niu,[1] Jingbo Liu,[4] Yuanfu Chen,[4] Kun-Lin Yang,[2] Xinhai Zhang,[5] Jong-Hyun Ahn,[3] and Hyunsoo Yang[1,6,*]

[1]*Department of Electrical Engineering and Computer Engineering, National University of Singapore, Singapore*
[2]*Department of Chemical and Biomolecular Engineering, National University of Singapore, Singapore*
[3]*School of Electric & Electronic Engineering, Yonsei University, Seoul, 120-749 and SKKU Advanced Institute of Nanotechnology, Sungkyunkwan University, Suwon, 440-746, Republic of Korea*
[4]*State Key Laboratory of Electronic Thin Films and Integrated Devices, University of Electronic Science and Technology of China, Chengdu 610054, China*
[5]*Institute of Materials Research and Engineering, 3 Research Link, 117602, Singapore*
[6]*Graphene Research Centre, National University of Singapore, 6 Science Drive 2, Singapore 117546, Singapore*
*\*eleyang@nus.edu.sg*

**Abstract:** Due to its high electrical conductivity and excellent transmittance at terahertz frequencies, graphene is a promising candidate as transparent electrodes for terahertz devices. We demonstrate a liquid crystal based terahertz phase shifter with the graphene films as transparent electrodes. The maximum phase shift is 10.8 degree and the saturation voltage is 5 V with a 50 μm liquid crystal cell. The transmittance at terahertz frequencies and electrical conductivity depending on the number of graphene layer are also investigated. The proposed phase shifter provides a continuous tunability, fully electrical controllability, and low DC voltage operation.

## 1. Introduction

Graphene has attracted much interest for a wide range of studies [1-3]. It is reported that the electrical mobility of single layer graphene can be as high as 250,000 $cm^2V^{-1}s^{-1}$ [1], which directly results in its high electrically conductivity. The transmittance of graphene is also outstanding at both visible [4] and terahertz (THz) frequencies [5], unlike the indium tin oxide (ITO) film which shows high transmittance at only visible frequencies [6], but very low at THz frequencies [7]. Together with the high chemical stability and mechanical strength [4], graphene is a promising candidate as transparent electrodes. Liquid crystal based devices with graphene as electrodes have been demonstrated at visible and near infrared frequencies [4], however graphene based phase shifters at THz frequencies have not been explored yet.

For the past three decades, much attention has been focused on THz frequencies due to its unique properties [8, 9]. Recently, graphene has been widely investigated at THz frequencies, such as graphene plasmonic structures [10-12], graphene modulators [13], and Dirac fermion dynamics measurements [14]. A phase shifter at THz frequencies is an important component for applications. Electrically or magnetically controlled phase shifters have been realized by controlling the liquid crystal alignment at room temperature. Due to the structural limitation, a high control voltage (> 100 V) is required for electrically controlled phase shift applications [15], and the magnetic field based phase shifters [16] are not ideal for integration due to the bulky magnets.

In typical THz phase shifter structures, the electric field is applied normal to the direction of THz beam propagation [15] because of the low transmittance at THz frequency of conventional transparent electrodes, which results in the decay in performance of devices. For example, ITO is widely used as transparent contacts at visible frequencies, however it is highly absorbing at THz frequencies (less than 10% transmission rates at 0.2 - 1.2 THz) [7].

By decreasing the thickness of the ITO films to enhance the transmittance at THz frequencies, the resistance of ITO film increases significantly [17]. A thin carbon nanotube film shows a low conductivity and poor transparency at THz frequencies [18, 19]. On the other hand, graphene shows excellent transmission in the THz range [5] as transparent electrodes without losing the electrical conductivity [20]. In addition, graphene enables the device to use direct current (DC) voltage which cannot be applied in ITO due to the diffusion of ions such as indium and tin [4].

In this study, a new THz phase shifter utilizing graphene and liquid crystals has been demonstrated at the frequency range of 0.2 to 1.2 THz. With chemical vapor deposition (CVD) graphene films as transparent electrodes, the new devices require low driving voltages (DC 5 V for saturation) for a 10.8° phase shift. Due to the pure electric control capability, the proposed phase shifters do not involve any magnet or any mechanical part, which makes it friendly for integration. In addition, our devices show a linear response at low bias voltage regime, which enables a continuous tunable operation.

## 2. Methods and experimental data

The graphene films utilized in this work were grown by CVD [21]. Single layer graphene was firstly grown on copper foils, and subsequently transferred onto quartz glass. The multilayer graphene films were prepared by stacking single layer films. Raman spectroscopy for single layer graphene shows the prominent G and 2D peaks without a detectable D peak, which indicates the good quality of graphene films as shown in Fig. 1(a) [22]. The inset of Fig. 1(a) shows the transport characteristic of a typical single layer graphene sample, indicating the graphene sample is p-type at zero bias voltage due to unintentional doping during copper etching process [23, 24]. With a graphene layer on top of the glass substrate, the transmittance reduces by ~2% at visible and near infrared spectrum range as shown in Fig. 1(b), which is comparable to the previous reports [25, 26].

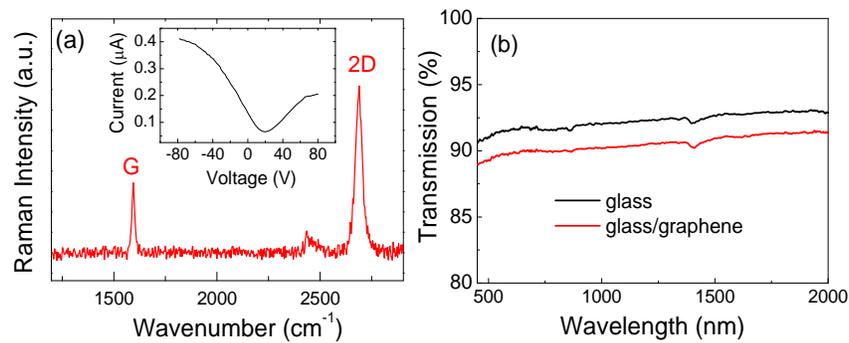

Fig. 1. (a) A Raman spectrum of single layer graphene. The inset of (a) is a current versus gate bias curve. (b) Transmittance of CVD graphene at visible and near infrared wavelengths.

A schematic diagram of the phase shifter is showed in Fig. 2(a). CVD graphene films were transferred onto the polished quartz glass substrates. Subsequently a thin Cr/Au ribbon (10 mm × 0.5 mm × 40 nm) was deposited onto one edge of each single layer graphene, which was used for applying bias voltages. For a uniform alignment of the liquid crystals, a layer of polyimide is required. Substrates with CVD graphene films were immersed in the polyimide solution for one hour and consequently dried. By gently rubbing the surface, the alignment of the polyimide layer was achieved for the alignment of liquid crystals. Next, two spacers of a thickness of 50 μm were used to separate the top and bottom structures, as shown in Fig. 2(a). Then the two glass substrates were sandwiched with the graphene layers facing each other. As the settling of the spacers, space of 8 mm × 8 mm × 50 μm was prepared for liquid crystals. In the last step, the liquid crystals 4'-n-pentyl-4-cyanobiphenyl (5CB) were filled into the prepared gap. In order to avoid unexpected electrical connections, two Cr/Au electrodes were mounted at different sides of the device structure with a slight misalignment as shown in Fig. 2(a). At room temperature (25 °C), the ordinary and extraordinary indices of refraction for 5CB are reported ~ 1.58 and 1.77 in the frequency range of 0.2-1.2 THz, respectively, and the birefringence value is 0.2 ± 0.02 [27]. The large and stable birefringence makes 5CB an ideal material for THz applications.

Due to the uniformly aligned polyimide, 5CB is initially aligned parallel to the substrate, as illustrated in Fig. 2(b). As the applied voltage between the two graphene electrodes increases, liquid crystals gradually switch to the perpendicular direction as shown in Fig. 2(c). Depending on the alignment angle of liquid crystals, the effective refractive index of 5CB for the THz pulse varies. The controllable transmission time delays are directly related to the tunable phase shift. Note that the initial alignment direction of 5CB without applied voltage needs to be parallel to the electrical field direction of the THz plane wave, which is also parallel to the film plane.

The measurements have been conducted in a commercial THz time domain spectroscopy setup (Teraview TPS spectra 3000) [27, 28], which can generate a well-defined plane wave pulse from 0.06 - 4.0 THz. For the time domain measurements, with any fixed bias voltage $v$, tens of measurements have been repeated and the time domain signal difference was less than 3 fs. The beam size is less than 5 mm in diameter, which is smaller than the area of liquid crystal cell (8 mm × 8 mm). The samples are located in a chamber with nitrogen and the temperature of the chamber is controlled at 25 ± 0.5 °C. A DC voltage is applied through the two Cr/Au electrodes shown in Fig. 2(a) labeled as (6). The voltage is gradually increased from 0 to 9 V. When applying the metallic or metal oxide films such as ITO as transparent electrodes, alternating current (AC) voltage is required to avoid the ion diffusion problem, which deteriorates the device performance and consequently lower the breakdown voltage. Since graphene films do not have the ion-injection issue, the DC voltage can be applied instead of AC voltage in our structure [4].

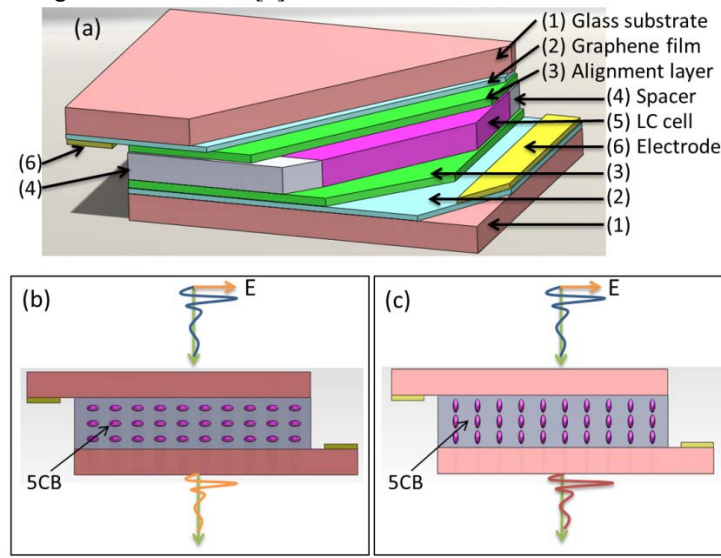

Fig. 2. (a) A schematic diagram of the graphene-based liquid crystal phase shifter at terahertz frequencies. The device consists of (1) glass substrates, (2) single layer CVD graphene films, (3) alignment layer (polyimide), (4) spacers (thickness 50 μm), (5) liquid crystal cell (5CB) (thickness 50 μm), and (6) Cr/Au electrodes (thickness 40 nm). (b) Before applying voltage, liquid crystals are aligned parallel to the film plane due to the polyimide layer. (c) With the bias voltage, liquid crystals are aligned perpendicular to the film plane. "E" denotes the electrical field direction of the THz pulses. The pulse transmission direction is perpendicular to the substrates.

In order to use CVD graphene films as transparent electrodes at THz frequencies, the transmittance of graphene at THz frequencies and its sheet resistance are very important factors. Figure 3(a) shows that, as the number of graphene layer increases, the pulse intensity of transmitted THz signal decreases. By comparing the peak value of the pulse, the transmittance of graphene on top of quartz glass are found to be 88%, 80.6%, 76.6%, and 73.8% for single, bi-, tri-, and 4-layer graphene films, respectively, as shown in Fig. 3(b). Interestingly, each additional layer of graphene contributes differently in THz absorption, unlike 2% absorption for each additional layer at visible and near infrared frequency range [25, 29]. In Fig. 3(c), the absorption rate, which is equal to the change of transmittance, at THz frequencies is plotted. As the number of layer increases, the amount of absorption is not linearly proportional to the number of graphene layer. The other important factor is the sheet resistance. By stacking the single layer films, the sheet resistance of the film reduces as shown in Fig. 3(d), implying a tradeoff between the conductivity and the transmittance at different number of graphene layer.

## 3. Results and discussion

The transmitted time domain waveforms of the THz pulses at different DC bias voltages are shown in Fig. 4(a), and the inset shows the signal between 190 to 230 ps. With increasing the bias voltage, the alignment of liquid crystals switches from the initial parallel direction to the perpendicular direction, with respect to the film plane. The perpendicular alignment of liquid crystal results in a smaller effective refractive index of 5CB, and subsequently induces a smaller time delay. The experimental data of phase shifts at 0.25, 0.5, 0.75, and 1 THz are shown in Fig. 4(b). The solid lines show the theoretical prediction which will be explained later.

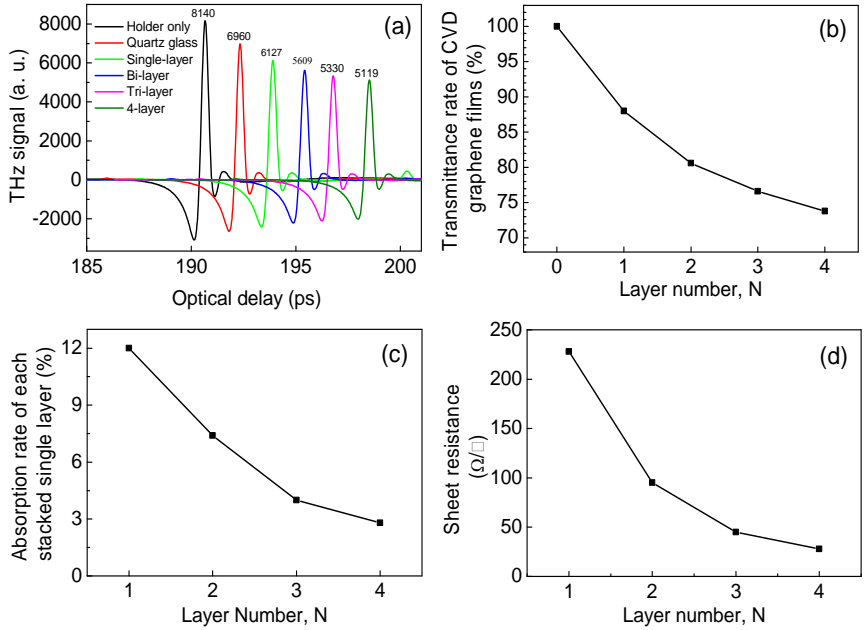

Fig. 3. (a) The THz pulse intensity through the sample holder, the bare glass substrate, single layer, bi-layer, tri-layer, and 4-layer graphene films. The each data set is shifted in time. (b) The transmittance of single layer, bi-layer, tri-layer, and 4-layer graphene films at THz frequencies. (c) The absorption rate of each additional stacked layer of graphene film. (d) Sheet resistance of CVD graphene films versus the number of stacked graphene layers.

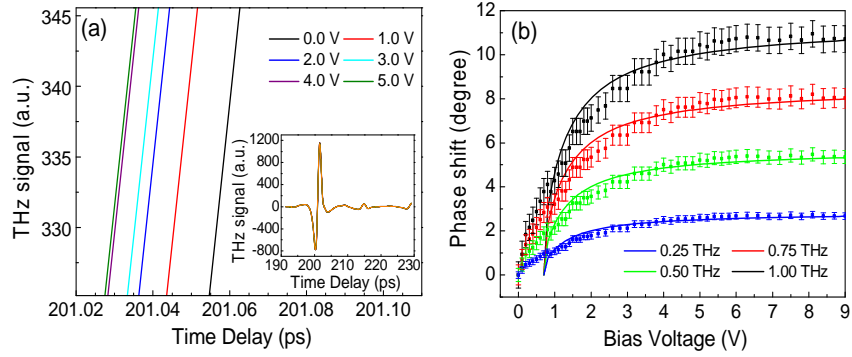

Fig. 4. (a) THz pulse signal transmitted through the phase shifter with different bias voltages. The inset shows the waveform of the THz signal in the full scanning range. Single layer graphene is used. (b) Voltage controlled phase shifts at different frequencies.

The threshold voltage (the Fréedericksz transition) of the device depends on the relationship

$$v_{th} = \pi \frac{L}{d} \sqrt{\frac{k_{11}}{\varepsilon_a \varepsilon_0}} \quad (1)$$

where $L$ is the distance between two electrodes, $d$ is the thickness of liquid crystal layer, $k_{11}$ is the splay elastic constant, $\varepsilon_a$ is the dielectric anisotropy, and $\varepsilon_0$ is the electric permittivity in free space [30]. Using the values of 5CB, $k_{11} = 5.24 \times 10^{-12}$ N, $\varepsilon_a = 11.92$, and $\varepsilon_0 = 8.85 \times 10^{-12}$ F/m, the threshold voltage $v_{th} = 0.7$ V can be calculated [31]. The phase shift is given by

$$\delta(v) = \frac{2\pi f d}{c} \Delta n_{eff}(v) \quad (2)$$

where $f$ is the frequency, $d$ is the thickness of the liquid crystal cell, $c$ is the speed of light in vacuum, and $\Delta n_{eff}(v)$ is the change of the effective refractive index [15]. In our structure $\Delta n_{eff}(v)$ is the function of applied voltage. Because of the thin cell utilized in this work, the uniformity of the electric field is not a concern and the position dependent phase shift is negligible. The field dependence of the birefringence for the voltage slightly larger than the threshold voltage can be described by

$$\Delta n_{eff}(v) = (n_e - n_o) \frac{n_o}{n_e} \left[ \left(1 + \frac{n_o}{n_e}\right)\left(\frac{v - v_{th}}{v_{th}}\right) \right] \quad (3)$$

where $n_o$ and $n_e$ are the ordinary and extraordinary refractive indices of 5CB [32]. For $v - v_{th} \gg v_{th}$,

$$\Delta n_{eff}(v) = (n_e - n_o) \frac{n_o}{n_e} \left[ 1 - \frac{2}{v} \sqrt{\frac{k_{11}}{\varepsilon_a \varepsilon_0}} \right] \quad (4)$$

is applied for predicting the value of $\Delta n_{eff}(v)$ [32, 33]. The values of $\Delta n_{eff}(v)$ at different voltages are used to plot the theoretical curves.

For $v > 0.7$ V, the experimental data are in good agreement with the theoretical prediction. Below the theoretical threshold voltage (0.7 V) there is a mismatch as shown in Fig. 4(b). Our devices show a very small threshold voltage, which leads to a linear phase shift operation with enhanced reliability in applications. In the above simple threshold equation, it is assumed that the coating layer is not affected by the electric field. However, there are some studies showing that if the coating layer can be oxidized/reduced by the electric field, the threshold voltage will decrease significantly [34]. The threshold voltage of the liquid crystal cell depends on anchoring energy from the alignment layer, the species of the liquid crystals, and the composition of the liquid crystals [30, 31]. In order to understand this discrepancy, further studies are required.

In previous THz phase shifters more than 100 V AC voltages were applied in order to saturate the device response due to a large separation of two electrodes [15]. However, in our cases only several volt of DC bias is enough to saturate the phase shift response. In addition, the desired phase shift can be tuned accurately in our structure due to the linear response around zero bias, whereas the linear response was not possible in previous reports [15]. The low threshold voltage and linear response at low DC bias voltages in our devices are advantageous in applications.

There are a few schemes to increase the maximum phase shifts in the proposed prototype phase shifters. One solution is to replace 5CB with other liquid crystals which have a larger birefringence [35]. Another method is to apply the piezoelectric materials, such as quartz crystals, to induce electrically controllable phase shifts which are expected to be significant in THz ranges. Furthermore, since the values of birefringence are relatively constant at THz frequency ranges [27], and the amount of phase shift is proportional to the operation frequency [15], the phase shift of our devices can easily reach 43.2° at 4 THz, and 90° at 8.3 THz.

## 4. Conclusion

We demonstrate a liquid crystal-based THz phase shifter with CVD graphene films as transparent electrodes, utilizing high transparency of graphene films at THz frequencies and the good electrical conductivity. It is found that the transmittance of graphene at THz frequencies shows a nonlinear characteristic with increasing the number of layer. The amount of phase shift at 1 THz gradually increases from 0° to 10.8° under the application of bias voltage. The proposed THz phase shifter is fully controlled electrically without involving any bulky magnetic component, implying easy integration into a small form factor. It provides a linear controllability of the amount of phase shift from very low bias voltages. A larger phase shift can be realized by replacing materials with larger birefringence or operating the devices at higher frequencies.